\documentclass[twocolumn,showpacs,preprintnumbers,amsmath,amssymb]{revtex4}

\def\bea{\begin{eqnarray}}
\def\eea{\end{eqnarray}}
\def\bfp{{\bf p}}
\def\simg{\,\hbox{\kern.1em \lower.6ex \hbox{$\sim$} \kern-1.12em
          \raise.6ex \hbox{$>$} }}
\def\bfq{{\bf q}}
\def\eq#1{(\ref{#1})}

\begin{document}

\title{Semiclassical description of shell effects in finite fermion systems}
\date{\today{}}
\author{Matthias Brack}
\affiliation{Institut f\"{u}r Theoretische Physik, Universit\"{a}t
Regensburg, D-93040 Regensburg, Germany}
\pacs{03.65.Sq}

\begin{abstract}
A short survey of the semiclassical periodic orbit theory, initiated 
by M. Gutzwiller and generalized by many other authors, is given. Via 
so-called semiclassical trace formmulae, gross-shell effects in bound 
fermion systems can be interpreted in terms of a few periodic orbits of 
the corresponding classical systems. In integrable systems, these are 
usually the shortest members of the most degenerate families or orbits, 
but in some systems also less degenerate orbits can determine the 
gross-shell structure. Applications to nuclei, metal clusters, 
semiconductor nanostructures, and trapped dilute atom gases are discussed.
\end{abstract}

\maketitle

\section{Introduction}
\label{secintr}

The Nilsson model \cite{sgn} gave rise to my first steps towards scientific
research. My teacher in theoretical physics was Kurt Alder, who had been a 
member of the theoretical 'Coulomb excitation crew' \cite{coulex} at the 
Niels Bohr Institute during the years when collective nuclear motion was 
explored and single-particle motion in deformed nuclei was studied. When 
I asked Alder about a possible subject for my diploma work, he gave me a 
copy of Sven G\"osta's famous paper and asked me to write a program for 
computing the deformed single-particle wavefunctions; he had become weary 
of interpolating their expansion coefficients in Nilsson's tables. I set 
about to diagonalize the Nilsson Hamiltonian using, however, not the 
spherical basis but that of the deformed harmonic oscillator, i.e., the 
asymptotic Nilsson states. The results were bound to be the same as those 
of Nilsson and were therefore not published \cite{mbdip}  -- but through 
this exercise I became initiated into the shell structure of nuclei. I felt 
very privileged when I later came to know the great human being behind the 
famous model.

In this paper I want to review the semiclassical description of shell-effects 
in finite fermion systems using the periodic orbit theory.
After a brief reminder about trace formulae, I will discuss some of their
applications to systems in four different branches of physics. Since all 
results have been published elsewhere, I will not reproduce here any 
figures, but just discuss in words some of the most important results and 
conclusions.

\section{Periodic orbit theory and gross-shell effects}
\label{secpot}

\subsection{Semiclassical trace formulae}
\label{sectrf}

The periodic orbit theory (POT) was initiated by M. Gutzwiller in a series 
of publications culminating in his seminal paper in 1971 that contains the 
semiclassical trace formula \cite{gutz}. It relates the quantum spectrum
$\{E_i\}$ (which we here assume to be discrete, although the inclusion of
a continuum is possible) of a hermitian Hamiltonian $\widehat H$ to the 
periodic orbits of the corresponding classical Hamiltonian $H(\bfq,\bfp)$.
The quantum-mechanical level density, defined as the sum of Dirac delta
functions peaked at the levels $E_i$, can be decomposed into a smooth
part ${\widetilde g}(E)$ and an oscillating part $\delta g(E)$:
\bea
g(E) = \sum_i \delta(E-E_i) = {\widetilde g}\,(E) + \delta g(E)\,.
\label{glev}
\eea
The smooth part contains by definition the average level density which
usually is a monotonously increasing function of $E$ and can be obtained in 
the extended Thomas-Fermi (ETF) model (see, e.g., \cite{book}, Ch.\ 4). The
oscillating part can be expressed by the {\it semiclassical trace formula}
\bea
\delta g_{sc}(E) \simeq \sum_{po} {\cal A}_{po}(E) 
                        \cos\,[S_{po}(E)/\hbar-\sigma_{po}\pi/2]\,.
\label{dgsc}
\eea
The sum is over all periodic orbits $(po)$ of the classical system, 
$S_{po}(E)=\oint\bfp\cdot d\bfq$ are their action integrals, the amplitudes
${\cal A}_{po}(E)$ depend on their stabilities and degeneracies, and
$\sigma_{po}$ are called the Maslov indices. The sum in \eq{dgsc} is an 
asymptotic one, correct to leading order in $1/\hbar$, and in non-integrable 
systems it is hampered by convergence problems \cite{chaos}. For isolated
orbits, Gutzwiller expressed \cite{gutz} the amplitudes ${\cal A}_{po}(E)$ 
in terms of their periods and stability matrices. 

The trace formula \eq{dgsc} was later generalized 
to billiard systems \cite{bablo} and to systems with continuous symmetries 
\cite{struma,bertab,crli}, including integrable systems. A relativistic 
trace formula for spin 1/2 particles was derived in \cite{boke}, and a 
nonrelativistic trace formula for particles with arbitrary spin $s$ in
\cite{plet}. In all cases, the trace formula has the same general form 
\eq{dgsc}, but the amplitudes ${\cal A}_{po}(E)$ take different forms. For 
isolated orbits, their $\hbar$ dependence is given by a factor $\hbar^{-1}$, 
while for orbits appearing in $f$-fold degenerate families, the amplitudes 
go like $\hbar^{-(1+f/2)}$.

In integrable and mixed-dynamical systems, periodic orbits can change their 
stability under the variation of a control parameter (e.g., the energy $E$, 
a potential parameter, or an additional external field) and thereby undergo 
bifurcations. In such situations, the amplitudes ${\cal A}_{po}$ diverge at the 
bifurcation points. The same happens also in limits where continuous symmetries 
are broken (or restored), since hereby the $\hbar$ dependence of the 
${\cal A}_{po}$ changes discontinuously. The remedy to remove these (unphysical!) 
divergences is to go beyond the stationary-phase approximation for the 
integration(s) used in the derivation of the semiclassical trace formula. This 
has, besides \cite{bablo,struma,bertab}, been developed most systematically in 
\cite{ozoha} for symmetry breaking and bifurcations, and in \cite{crpert} for 
symmetry breaking in weakly perturbed integrable systems, leading in all cases 
to local uniform approximations with finite amplitudes ${\cal A}_{po}$. Global 
uniform approximations which yield finite amplitudes at symmetry-breaking and 
bifurcation points, and far from them go over into the standard (extended) 
Gutzwiller trace formula, were developed for the breaking of U(1) symmetry in 
\cite{toms}, for some cases of U(2) and SO(3) symmetry breaking in \cite{hhuni}, 
for the symmetry breaking U(3) $\to$ SO(3) in \cite{boys}, and for various 
types of bifurcations in \cite{ssun}. (Details and further references may be 
found in \cite{book}, Ch.\ 6.3.)

For interacting finite fermion systems described in the mean-field approximation
(i.e., in Hartree-Fock or density functional theory), one can also obtain 
semiclassical trace formulae for the oscillating parts of the total binding
energy $E_b$ and the particle number $N$. Hereby one writes, similarly to 
\eq{glev}, $E_b={\widetilde E}_b+\delta E$ and $N={\widetilde N}+
\delta N$. The average quantities ${\widetilde E}_b$ and ${\widetilde N}$ are
taken from the ETF model, and for the oscillating parts one finds 
\cite{book,struma}
\bea
\delta E_{sc} \!\! & \simeq & \! \sum_{po} {\cal A}_{po}(\lambda)
                         \left(\frac{\hbar\,}{T_{po}}\!\right)^{\!2}\!
                         \cos\!\left[\frac{S_{po}(\lambda)}{\hbar}
                         -\sigma_{po}\frac{\pi}{2}\right]\!,
                         \nonumber\\
\delta N_{sc} \!\! & \simeq & \! - \! \sum_{po} {\cal A}_{po}(\lambda) 
                         \left(\frac{\hbar\,}{T_{po}}\!\right)
                         \sin\!\left[\frac{S_{po}(\lambda)}{\hbar}
                         -\sigma_{po}\frac{\pi}{2}\right]\!,~~
\label{dgEN}
\eea
both to be evaluated at the Fermi energy $\lambda(N)$ for a given number of
particles $N$. The periodic orbits are ideally those of the classical 
counterpart of the self-consistent mean field, which for practical 
purposes often is taken as a shell-model type potential. 

\subsection{Coarse-graining and finite temperatures}
\label{seccoa}

Our present emphasis in the use of POT is not the full quantization of the
spectra of finite fermion systems, but on the semiclassical description of 
their {\it gross-shell structure}. For this purpose we {\it coarse-grain} 
the quantum spectrum by a convolution of the level density \eq{glev} with a 
normalized Gaussian of width $\gamma$:
\bea
g_{qm}(E,\gamma) = \frac{1}{\gamma\sqrt{\pi}} \sum_i e^{-(E-E_i)^2\!/\gamma^2}.
\label{ggam}
\eea
The coarse graining of the trace formula \eq{dgsc} gives, using the
stationary-phase approximation for the convolution integral, an extra exponential
factor in the trace formula:
\bea
\delta g_{sc}(E,\gamma) & \simeq & \sum_{po} {\cal A}_{po}(E)\, 
                                   e^{-(\gamma T_{po}/2\hbar)^2}\times
                                   \nonumber\\ 
                        & &        ~~~~~~~~~~~~~~~
                                   \cos\!\left[\frac{S_{po}(\lambda)}{\hbar}
                                   -\sigma_{po}\frac{\pi}{2}\right]\!.
\label{dgscg}
\eea
The same exponential factor appears in the trace formulae \eq{dgEN} for 
$\delta E_{sc}(\lambda,\gamma)$ and $\delta N_{sc}(\lambda,\gamma)$. It
suppresses the contributions from orbits with larger periods $T_{po}$. A 
similar suppression of longer orbits and the overall amplitude of the shell 
effects occurs at {\it finite temperatures}. E.g.\ in a grand-canonical system 
at temperature $T$, the oscillating part of the free Helmholtz energy has 
the trace formula
\bea
\delta F_{sc}(\lambda,T) & \simeq & \sum_{po} {\cal A}_{po}(\lambda) 
                                    \left(\frac{\hbar\,}{T_{po}}\!\right)^{\!2}\!
                                    \frac{\tau_{po}}{\rm{Sinh}(\tau_{po})}
                                    \times\nonumber\\ 
                         & &        ~~~~~~~~~~~~~~~
                                    \cos\!\left[\frac{S_{po}(\lambda)}{\hbar}
                                    -\sigma_{po}\frac{\pi}{2}\right]\!.
\label{dFsc}
\eea
Here $\tau_{po}=k_BT\pi T_{po}(\lambda)/\hbar$ and $k_B$ is the Boltzmann
contant. In both situations the {\it gross-shell effects} are dominated by the 
{\it shortest periodic orbits} of the system \cite{struma}.

In mixed-dynamical and integrable systems, orbits with different degrees $f$ 
of degeneracy can coexist. The gross-shell structure then results from a
competition between the periods and the degeneracies of the shortest orbits
\cite{ozotom}. Whereas their coarse-grained semiclassical amplitudes decrease 
with growing period $T_{po}$, they increase with growing degeneracy $f$ due to 
their dependence $\propto\hbar^{-(1+f/2)}$ already mentioned above. 

In arbitrary {\it spherical three-dimensional systems}, the most degenerate orbits
undergo both radial and angular oscillations with rational frequency ratios (cf.\ 
\cite{bertab,boys,bm}):
\bea
\omega_r\,:\,\omega_\phi = n:m\,, \qquad |n|,|m|\in\mathbb{N}\,.
\label{reson}
\eea
(For physical reasons, only pairs of integers $n,m$ with equal signs are allowed;
orbits with negative $n,m$ correspond to the time-reversed of the orbits with positive
$n,m$). The orientiations of these 'rational tori' can be rotated about three Euler 
angles without changing their shapes, periods or actions; therefore they appear in
three-fold degenerate families ($f\!=3$). Their existence for arbitrary ratios 
$n\!:\!m$ depends, however, on the form of the radial potential $V(r)$ (cf.\ 
\cite{boys,arita}) and, in general, on the energy $E$. The special orbits with angular 
momentum $L\!=\!0$ and $L\!=\!L_{max}$, corresponding to librating 'diameter' and 
rotating 'circle' orbits, 
respetively, form families with only two-fold degeneracy ($f\!=2$), since one of the 
three Euler rotations does not change their orientations. The {\it Coulomb potential}
$V(r)=-\alpha/r$ has an extra dynamical symmetry, leading to O(4); here all orbits
have $n\!:\!m=1\!:\!1$. {\it Spherical harmonic oscillators} $V(r)=ar^2$ also have an 
extra dynamical symmetry, leading in three dimensions to U(3); here all orbits have 
$n\!:\!m=2\!:\!1$. In each of these two special potentials, {\it all} periodic orbits 
(including the diameters and circles) form one family with degeneracy $f\!=4$. 
Quantum-mechanically, the dynamical symmetries reflect themselves in an accidental 
extra degeneracy of the eigenvalue spectrum $\{E_i\}$. The semiclassical trace formulae 
for $\delta g(E)$ of these systems, added to their ETF expressions for ${\widetilde 
g}\,(E)$, reproduce the {\it exact} quantum-mechanical level densities according to 
\eq{glev} [see \cite{book}, Eqs.\ (3.144) and (3.69) for their explicit analytical 
expressions].

\section{POT for finite fermion systems}
\label{secapp}

\subsection{Nuclei}
\label{secnuc}

{\it 1. Ground-state deformations.}
Strutinsky {\it et al.}\ \cite{struma,strdo} were the first to extend the POT to 
systems with continuous symmetries and to apply it to the study of gross-shell effects 
in nuclei. In \cite{strdo} they studied the periodic orbits in a spheroidal cavity 
with axis ratio $\eta$ as a model for the mean field of a deformed nucleus. They
plotted the shell-correction energy $\delta E$, obtained quantum-mechanically using 
Strutinsky's shell-correction method \cite{strut} with the spectra of realistic 
deformed Woods-Saxon potentials \cite{fuhil} with the same spheroidal deformations 
(including spin-orbit interaction),  versus particle number $N$ and deformation $\eta$. 
The slopes of the valleys in these deformation ener\-gy surfaces $\delta E(N,\eta)$ 
corresponding to the ground-state deformations could then be correctly reproduced by 
the condition that the actions $S_{po}$ of the shortest and most degenerate periodic 
orbits in the spheroidal cavity (with $f\!=2$) \cite{sphero} be constant. Contributions 
of orbits with $f\!=1$ were negligible. (The Fermi energies corresponding to the 
spherical magic numbers had to be adjusted, as no spin-orbit interaction was included 
in the cavity model.) Although this was only a qualitative result, it proved the 
correctness of the concept to interpret quantum-mech\-anical gross-shell effects 
semiclassically in terms of short periodic orbits of the corresponding classical system.
The use of a cavity with infinitely steep walls for the mean field hereby justifies 
itself through the short range and the saturating property of the effective 
nucleon-nucleon interaction, leading to steep walls of the self-consistent 
Hartree-Fock potentials or their approximations by Woods-Saxon type shell-model 
potentials.

\vspace*{0.07cm}

{\it 2. Left-right asymmetry of fission barriers.}
A prominent manifestation of shell effects is the 'dou\-ble-humped' fission barrier of
nuclei in the actinide region \cite{sven}. One particular aspect is that of the onset
of a left-right asymmetry of the fissioning nuclear shapes which eventually leads to 
the asymmetric mass distributions of the fission fragments. Since Sven G\"osta Nilsson
and his group, and other scientists in Lund, were much involved in the study of this 
shell effect, I may dwell a little on its history.\\ 
The mixing of pairs of single-particle states with opposite parities in a spheroidal 
harmonic-oscillator potential was studied earlier \cite{leeing} as a possible 
mechanism leading to 'pear-shaped' nuclei. In 1962, S. A. E. Johansson \cite{johan} 
took this question up and investigated the possibility of octupole-deformed fission 
barriers. Since Strutin\-sky's shell-correction method \cite{strut} did not yet exist 
at that time, no realistic fission barriers could be obtained qantum-mechanically with 
the Nilsson model. Johansson showed, however, that at the typical deformations of the 
actinide fission barriers predicted by the liquid-drop model (LDM) \cite{ldm}, the 
mixing of single-particle orbits of the type used in \cite{leeing} leads to an 
instability of the barrier against octopole shapes. 
Using the shell-correction method with the Nilsson 
model, P. M\"oller and S. G. Nilsson \cite{asymn} obtained in 1970 the instability of 
the outer fission barrier against a suitable mixture of $\epsilon_3$ and $\epsilon_5$ 
deformations. Thus, the onset of the fission mass asymmetry was clearly a 
quantum-mechanical shell effect that could not be explained by the classical LDM model. 
In a detailed microscopical study, C. Gustafsson, P. M\"oller, and S. G. Nilsson 
\cite{gumni} showed a year later that those pairs of single-particle states, which are 
most sensitive to the left-right asymmetric shapes and hence responsible for their 
onset, have their wavefunction nodes and extrema on parallel planes at and near the 
waist-line of the fissioning nucleus perpendicular to its symmetry axis.\\ 30 years 
later, in a Lund-Regensburg-Dresden collaboration \cite{chaofi}, this effect was 
studied semiclassically. The POT had been used in the same collaboration \cite{brs} for 
cavities with the $(c,h,\alpha)$ shapes of \cite{fuhil}, for which the mass asymmetry 
of the outer fission barrier of actinide nuclei had also been obtained 
quantum-mechanically in \cite{asyplb}. The shortest periodic orbits here are families 
with $f\!=1$, having the axial U(1) symmetry; they are simply the diagonal, triangular 
and square-shaped orbits in the circular planes perpendicular to the nuclear symmetry 
axis. The semiclassical trace formula (with a uniform approximation for the bifurcation 
occurring at the onset of the neck, where the orbits in the central plane become 
unstable and give birth to two new parallel planes with stable orbits) was shown 
\cite{brs} to yield realistic deformation enery surfaces $\delta E_{sc}(c,h,\alpha)$ 
in the region of the outer fission barrier, predicting its instability against the 
asymmetry parameter $\alpha$ in good agreement with the old quantum-mechanical 
results \cite{asyplb}. (Again, the spin-orbit interaction was omitted; the 
Fermi energy as the only parameter was adjusted to yield the fission isomer minimum at 
the correct deformation.) Similarly to \cite{strdo}, the valley of steepest descent 
through the deformation energy surface, leading over a left-right asymmetric outer 
saddle, is obtained by the stationary condition $\delta S_{po}=0$ for the shortest 
orbits. In \cite{chaofi}, it was shown that these orbits are situated in a very small 
regular island of a dominantly chaotic phase space. An approximate 
Einstein-Brillouin-Keller (EBK) quantization of the linearized classical motion 
in these regular islands reproduced rather precisely the quantum-mechanical energies of 
those diabatic single-particle states with opposite parity which are most sensitive to 
the $\alpha$ deformations and hence quantum-mechanically responsible for the $\alpha$ 
instability of the outer barrier. Furthermore, their wavefunctions were found to have 
their nodes and extrema precisely in the planes near the nuclear waist-line that 
contain the shortest periodic orbits responsible semiclassically for the asymmetry 
effect.\\
This application of the POT represents an interesting example for the 
classical-to-quantum correspondence of the interplay between chaos and order: a tiny 
regular island in an almost chaotic phase space causes a quantum shell effect in an 
interacting many-body system with observable consequences in the form of the mass 
asymmetry of the fission fragments.

\subsection{Metal clusters}
\label{secclu}

\vspace*{-0.25cm}

Metal clusters are interesting finite fermion systems which allow one to study the
transition from atoms over molecules towards condensed matter \cite{deheer}. 
In the simplest theoretical description, the so-called 'jellium 
model', the ions are replaced by a structureless but deformable positive background 
and the systems of $N$ interacting valence electrons in the external jellium potential 
are studied \cite{braclu}. 
In neutral clusters, the eletrostatic long-range forces cancel and the valence
electrons are only bound by the short-ranged exchange and correlation effects. Neutral 
metal clusters therefore have much in common with nuclei. One difference to nuclei is 
that there is no measurable spin-orbit interaction in most metal clusters. The magic 
numbers $N_i$ of the smallest spherically stable clusters correspond to those of a 
harmonic oscillator ($N_i=2,8,20,40$) \cite{deheer}. For the analysis of early 
experimental abundance spectra of small sodium clusters, the Nilsson model without 
spin-orbit term was successfully employed \cite{clem} to interpret the regions between 
the spherical shell closures in terms of prolate and oblate deformations. In a 
self-consistent mean-field description \cite{eka1,genz}, the average potential of 
clusters with $N\simg 80$ valence electrons has steep walls like heavy nuclei, and 
therefore cavity models provide again a good approximation. 

One early result of POT was the observation by Balian and Bloch \cite{bablo} that the
coarse-grained level density of a {\it spherical cavity} exhibits a pronounced {\it 
beating pattern:} a rapid regular oscillation, reflecting the shell structure of the 
spectrum, modulated by a slow oscillation reaching over some 13 - 14 shells. From the 
trace formula derived in \cite{bablo} one sees that the beat comes about by the 
interference of the shortest periodic orbits of highest degeneracy ($f\!=3$), which 
here are the triangle ($n\!:\!m=3\!:\!1$) and square ($4\!:\!1$) orbits. (The diameter 
orbit with degeneracy $f\!=2$ can be neglected.) The rapid shell oscillations are 
determined by the average length of these two orbits, while the period of their 
amplitude modulation, the so-called `super-shell' oscillation, is given by the
difference of their lengths. 

The numbers of fermions needed to reach the first super-shell node is of the order of
$\sim 800 - 1000$. Super-shells can therefore not be seen in nuclei. Neutral metal
clusters, however, can be made arbitrarily large. This inspired Nishioka {\it et al.}\
\cite{nishi} to study the super-shell structure in Woods-Saxon potentials fitted to 
self-consistent mean fields \cite{eka1,bra89}, and to predict that it should be 
observable
in metal clusters. Indeed, the super-shells were experimentally observed, for the first 
time in supersonic beams of hot sodium clusters \cite{klavs}. Their abundance in an 
adiabatically expanding beam is dominated by their stability against evaporation of 
single atoms and exhibits pronounced peaks at the spherically magic numbers. The larger 
ones, $N_i=92, 138, 192, 264, \dots$, are almost exactly those of a spherical cavity. 
By a suitable extraction of the oscillating part of the abundance spectra, the 
super-shell beat could clearly be exhibited \cite{klavs}, with its first node appearing 
around $N\sim 900$.

Plotting the cube roots of the magic numbers, $N_i^{1/3}$ which are proportional to 
the r.m.s.\ radii of the magic clusters, against the shell number $i$, one obtains a 
straight line with a slope $s=N_{i+1}^{1/3}-N_i^{1/3}$. The experimental value of this 
slope is $s_{exp} = 0.61\pm 0.01$; it was confirmed by later experiments, also with 
other types of metal clusters (their different Wigner-Seitz radii do not affect the 
value of $s$, as long as they are single-valenced) \cite{deheer,braclu}. Around 
$i\simeq 14$, there is a slight discontinuity in the plot $s(i)$ before it continues 
again with the same slope. This is due to the phase change of the rapid shell 
oscillations when passing through the first super-shell node. The value of this slope 
predicted by the POT of \cite{bablo,nishi} is $s_{pot}=0.603$, that of the 
self-consistent quantum-mechanical mean-field calculations is $s_{mf}=0.61$, both in 
perfect quantitative agreement with experiment.

This provides another example of the good agreement of POT with both experiment and
quantum mechanics. An easily readable account of the super-shells in metal clusters is 
given in \cite{sciam}.

Unfortunately, it is very difficult to extract spectroscopic data on the electronic
single-particle states in metallic clusters. The most direct access to their shapes is
given by the so-called 'Mie plasmons', the collective dipole oscillations of the valence
electrons against the ions \cite{eka2,bra89} (which are the origin of the colors in 
stained-glass windows \cite{krvol}). Similarly to the giant-dipole resonances in nuclei, 
their splitting gives evidence of the average deformations of the clusters. This gave 
rise to a series of theoretical investigations of cluster deformations using deformed 
shell models (see \cite{braclu}). For applications of the POT one employs hereby most 
comfortably the spheroidal cavity model used also in \cite{struma}. I refer to 
\cite{frauen} for a detailled account 
containing also a nice application of the POT to the semiclassical interpretation of 
moments of inertia. The effects of weak magnetic fields on spherical metal clusters 
were studied in \cite{kaoclu}. In \cite{pash}, spheroidal cavities with the lowest 
multipole deformations $\epsilon_2$, $\epsilon_3$, and $\epsilon_4$ were studied, and 
the perturbative trace formula of \cite{crpert} (cf.\ also \cite{peter}) was used to 
predict their ground-state deformations. The results were in very good agreement with 
those of quantum-mechanical shell-correction calculations using the spectra of the 
same cavities.

\vspace*{-0.2cm}

\subsection{Semiconductor nanostructures}
\label{secnan}

\vspace*{-0.2cm}

Semiconductor heterostructures can be used to construct two-dimensional systems of
quasi-free electrons on the nanometer scale. With the help of external metallic gates 
or lithography, the electrons can further be laterally confined to form so-called 
quantum dots, quantum channels, quantum wires, antidot superlattices, etc.\ 
\cite{nano,reimat}. Applying a perpendicular magnetic field $B$, one can measure the 
magneto-resistance of such devices. Under suitable experimental circumstances, both 
the mean free path and the phase coherence length can be made larger than the sizes of 
these structures, so that quantum interference still takes place while the dimensions 
are large enough to allow for a semiclassical description. Nanostructues are therefore 
ideal tools to study the interplay between classical and quantum mechanics.

Weiss {\it et al.}\ \cite{weiss} measured the resistance of andidot superlattices and 
found oscillations which can be explained classically by the commensurability of 
cyclotron orbits with the superdot lattice: when an electron is trapped in a cyclotron 
orbit that fits around 1, 2, 4, 9, etc.\ antidots, it does not contribute to the 
conductance and hence a peak is seen in the magneto-resistance (see \cite{weri} for 
an easily readable account). In weak $B$ fields at very low temperatures, some rapid 
$B$-periodic oscillations could be observed. They could be interpreted semiclassically 
by the interference of different trapped periodic orbits of comparable lenghts; the 
linear response of the system to the $B$ field was hereby described by a semiclassical 
version of the Kubo theory yielding a trace formula for the conductance \cite{kubosc}. 
These so-called 'Aharonov-Bohm (AB) oscillations' can be easily understood in a 
perturbative approach \cite{crpert} in which the effect of the magnetic field is 
taken into account to lowest order only in the actions $S_{po}=\oint\bfp\cdot d\bfq$, 
while the shapes of the orbits are left unchanged. Under the canonical substitution 
$\bfp\to\bfp-(e/c){\bf A}$, where ${\bf A}$ is the vector potential with ${\bf B}\!=
\!\nabla\!\times\!{\bf A}$, the action of a periodic orbit changes like
\bea
S_{po}\to S_{po}-(e/c)\oint {\bf A}\cdot d\bfq=S_{po}-(e/c)\Phi_{po}\,,
\label{actab}
\eea
where $\Phi_{po}=\oint {\bf B}\!\cdot\!d{\bf F}_{po}$ is the magnetic flux through the 
{\it area} $F_{po}$ surrounded by the orbit. Consequently, the perturbed semiclassical
trace formula is modified only by a factor $\cos\,(e\Phi_{po}/\hbar c)$ containing the
Aharonov-Bohm phase which causes the $B$-periodic oscillations.

Similar AB oscillations have also been measured in the magneto-conductance of a circular 
quantum dot containing some $\sim 1200$ to $2000$ electrons \cite{pelrei}. They could 
be qualitatively well explained \cite{qdot} by the perturbed level density $\delta 
g(E,B)$ of a two-dimensional circular billiard. This is an integrable system whose 
trace formula is analytically known \cite{disk}. The conductance oscillations as 
functions of the radius of the quantum dot (regulated experimentally by the applied 
gate voltage) were well reproduced by the average length of the shortest orbits (here: 
diameters and triangles), while the period of the AB oscillations according to 
\eq{actab} was well reproduced by the area enclosed by the triangular orbit \cite{qdot}.

An analytical trace formula for the circle billiard in arbitrarily strong transverse
magnetic fields $B$ was given in \cite{blaqd}, and the magnetization of quantum dots
was studied semiclassically in \cite{kaoqd}.

In a mesoscopic semiconductor channel with two antidots, the magneto-conductance was 
measured in \cite{goul} and also found to exhibit AB oscillations. These could be well 
explained \cite{jphd} using the semiclassical Kubo formula \cite{kubosc} with a suitably 
modeled two-dimensional confinement potential for the channel (with antidots), which 
represents a mixed-dynamical system with a rather chaotic phase space. Plotting the 
maxima of the experimental AB oscillations versus magnetic field $B$ and gate voltage 
(which regulates the radii of the antidots), one obtains a grid of lines exhibiting some 
characteristic displacements \cite{goul}. At first sight, these might be attributed to 
missing flux units. Quantum-mechanical calculations \cite{kirc} reproduced these 
displacements but could not explain the physics behind them. In the 
semiclassical calculations the displacements could, in fact, be attributed to 
bifurcations of some of the trapped periodic orbits \cite{jphd,chan}.

\vspace*{-0.35cm}

\subsection{Trapped dilute atomic gases}
\label{sectrp}

\vspace*{-0.2cm}

I mention only briefly the finite fermion systems produced by confining diluted
fermionic atom gases, e.g.\ in magneto-optic traps \cite{grimm}. Hartree-Fock 
calculations for $N$ harmonically trapped atoms with a short-ranged repulsive 
interaction \cite{yylund} yielded shell effects $\delta E(N)$ in their total binding 
energies $E_b(N)$ which remind about the super-shells discussed above. More details are 
given in \cite{magnus}; let me just emphasize here that the origin of the beating shell 
structure is different here from that in a spherical cavity \cite{bablo}. The 
self-consistent mean field can be modeled by a perturbed harmonic oscillator 
$V(r)=ar^2+\epsilon\,r^4$. For such potentials it was shown recently \cite{boys} that 
the gross-shell structure is dominated by the two-fold degenerate diameter and circle 
orbits, whose interference explains the super-shells found in \cite{yylund}. The 
shortest three-fold degenerate orbits have frequency ratios $n\!:\!m\geq 7\!:\!3$ and 
contribute only to finer details of the quantum spectrum at relatively high energies.

\end{document}